\documentclass[11pt]{article}
\usepackage{BSMvietnam,epsfig}

\def\Journal#1#2#3#4{{#1} {\bf #2}, #3 (#4)}

\def\NPB{{\em Nucl. Phys.} B}
\def\PLB{{\em Phys. Lett.}  B}
\def\PRL{\em Phys. Rev. Lett.}
\def\PRD{{\em Phys. Rev.} D}

\def\ra{\rightarrow}

\def\be{\begin{equation}}
\def\ee{\end{equation}}
\def\bea{\begin{eqnarray}}
\def\eea{\end{eqnarray}}

\def\code{{\tt SUSY\_FLAVOR}}

\def\webpage{{\tt http://www.fuw.edu.pl/susy\_flavor}}

\newcommand{\bsmumu}{B_s^0 \rightarrow \mu^+ \mu^{-}}
\newcommand{\cbsmumu}{\mathcal{B}(B_s^0 \rightarrow \mu^+ \mu^{-})}
\newcommand{\bs}{B_s^0 \rightarrow \mu^+ \mu^{-}}

\newcommand{\lsim}{\raisebox{-0.13cm}{~\shortstack{$<$\\[-0.07cm] $\sim$}}~}

\begin{document}

\vspace*{3cm} \title {{\tt SUSY\_FLAVOR} library and constraints on
  $\bsmumu$ decay rate~\footnote{to appear in proceedings of the
    Rencontres du Vietnam ``Beyond The Standard Model of Particle
    Physics'', Qui Nhon, Vietnam July 15-21, 2012}}

\author{J.  ROSIEK}

\address{Institute of Theoretical Physics, Physics Department,
  University of Warsaw, Poland}

\maketitle\abstracts{I present \code{} - a Fortran 77 program able to
  calculate simultaneously 29 low-energy flavor and CP-violating
  observables in the general $R$-parity conserving MSSM, including the
  case of large flavor violation in the sfermion sector.  \code{} v2
  performs also the resummation of chirally enhanced corrections,
  arising in large $\tan\beta$ regime and/or large trilinear soft
  mixing terms, to all orders of perturbation theory.  I discuss an
  example of application of \code{} to analysis of the $\bsmumu$ decay
  rate in the MSSM.}

\section{Introduction}

Flavor physics was in the recent years one of the most active fields
in the high energy physics.  New experiments helped to improve the
accuracy of various measurements related to rare decays and put strong
constraints on the flavor structure of physics beyond the Standard
Model (SM), in particular imposing stringent limits on the flavor- and
CP- violating parameters of the Minimal Supersymmetric Standard Model
(MSSM).  It is then increasingly important to have an universal
computational tool which helps to compare new data with the
predictions of the MSSM.  Developing such a tool is a non-trivial task
requiring extensive and tedious calculations.  Numerous existing
analyses usually consider only a few rare decays simultaneously and in
most cases they are restricted to the case of so-called Minimal Flavor
Violation (MFV) scenario, where the CKM matrix is the only source of
CP and flavour violation~\cite{MFV}.

Based on a series of papers where many rare processes were analyzed
within the general MSSM~\cite{MIPORO}$^-$\cite{BEJR}, a library of
relevant computer codes has been published as \code{} v1~\cite{SFLAV}.
\code{} v2~\cite{SFLAV2} is in addition capable of resumming leading
chirally enhanced corrections, important in the regime of large
$\tan\beta$ or large trilinear $A$-terms~\cite{LTB,JRCRIV}, to all
orders of perturbation theory and for any pattern of sfermion mass
matrices - a unique feature not shared by other publicly available
programs.  In this article I briefly summarize the main features of
\code{} and present example of its application to estimating the
$\bsmumu$ decay rate in the MSSM.

\section{\code{} structure, input parameters and calculated observables}
\label{sec:init}

\code{} is capable of calculating physical observables within the most
general $R$-parity conserving MSSM, with one exception: currently it
assumes massless neutrinos (and no right neutrino/sneutrino
fields~\cite{DEHARO}), so the PMNS mixing matrix does not appear in
the couplings.

\code{} has been in development long before the Les Houches
Accord~\cite{SLHA} (SLHA) for common MSSM conventions was agreed on.
Thus, the internal routines of the library follow the conventions of
earlier paper~\cite{PRD41}.  However, by default \code{} can be
initialized with a SLHA2 compatible set of parameters - all
translations are done internally.  Note that in \code{} one can also
use so-called ``non-analytic $A$-terms'' of the form $A_l^{'IJ}
H^{2\star} L^I E^J + A_d^{'IJ} H^{2\star} Q^I D^J + A_u^{'IJ}
H^{1\star} Q^I U^J$, which are not included in the standard SLHA2
parametrization.

\begin{table}[t]
\caption{One loop parton level formfactors implemented in \code.
  $I,J,K,L$ denote flavor indices.}
\begin{center}
{\small
\label{tab:green}
\begin{tabular}{|cp{1mm}|p{1mm}cp{1mm}|p{1mm}c|}
\hline
Box &&& Penguin &&& Self energy \\ 
\hline 
$d^Id^Jd^Kd^L$ &&& $Z\bar d^I d^J$, $\gamma \bar d^I d^J$, $g \bar d^I
d^J$ &&& $d^I$-quark \\
$u^Iu^Ju^Ku^L$ &&& $H_i^0 \bar d^I d^J$, $A_i^0 \bar d^I d^J$ &&&
$u^I$-quark \\
$d^Id^Jl^Kl^L$ &&& $H_i^0 \bar u^I u^J$, $A_i^0 \bar u^I u^J$ &&&
charged lepton $l^I$\\
$d^Id^J\nu^K\nu^L$ &&& $\gamma\bar l^I l^J$ &&& \\ 
\hline
\end{tabular}
}
\end{center}
\end{table}

Calculations in \code{} take the following steps~\cite{SFLAV2}:

\noindent {\em 1.  Parameter initialization.}  Users can adjust the
basic SM parameters and initialize all (or the chosen subset of) Higgs
sector parameters and supersymmetric soft masses and couplings (which
must be specified at the SUSY scale).

\noindent {\em 2.  Calculation of the physical masses and the mixing
  angles.}  In the next stage, the eigenvalues of the mass matrices of
all MSSM particles and their mixing matrices at tree level are
calculated numerically, without any approximations.

\noindent {\em 3.  Resummation of the chirally enhanced effects.} In
the regime of large $\tan\beta$ and/or large trilinear SUSY breaking
terms, large chirally enhanced corrections to Yukawa couplings and CKM
matrix elements arise~\cite{LTB}.  They are resummed to all orders of
perturbation theory using the formalism developed in~\cite{JRCRIV}.
The level of resummation (no resummation, approximate analytical
resummation in the decoupling limit, iterative numerical resummation)
is a user defined option.

\noindent {\em 4.  Calculation of the Wilson coefficients at the SUSY
  scale}.  In the current version, \code{} calculates Wilson
coefficients generated by the diagrams listed in Table~\ref{tab:green}
(routines for given formfactor accept fermion generation indices as
input parameters).

\noindent {\em 5.  Strong corrections.}  After evaluating virtual SUSY
contributions, \code{} performs the QCD evolution of the Wilson
coefficients from the high (SUSY or top quark mass) scale to the low
energy scale appropriate for a given decay.  Necessary hadronic matrix
element estimates and other QCD related quantities are treated as
external parameters, initialized to the default values extracted from
analyses done within the SM but also directly modifiable by users.

\noindent {\em 6.  Evaluation of physical observables.}  Finally
physical observables are calculated and printed out.   Current list of
processes implemented in \code{} v2 is listed in Table~\ref{tab:proc}.

\begin{table}[htb]
\caption{List of observables calculated by \code{} v2 and their
  currently measured values.
}
\begin{center}
{\small
\label{tab:proc}
\begin{tabular}{|lr|lr|}
\hline
Observable &Experiment &  Observable &Experiment \\ 
\hline

\multicolumn{2}{|c|}{$\Delta F=1$} & \multicolumn{2}{|c|}{$\Delta
  F=0$}\\

\hline

$\mathcal{B}(\mu\to e \gamma)$ & $<2.8 \times 10^{-11}$ &
 $\frac{1}{2}(g-2)_e$ &$(1 159 652 188.4 \pm4.3) \times 10^{-12}$ \\

$\mathcal{B}(\tau\to e \gamma)$ & $<3.3\times 10^{-8}$ &
 $\frac{1}{2}(g-2)_\mu$ &$(11659208.7\pm8.7)\times10^{-10}$ \\

$\mathcal{B}(\tau\to \mu \gamma)$ & $<4.4\times 10^{-8}$ &
 $\frac{1}{2}(g-2)_\tau$ &$<1.1\times 10^{-3}$ \\

$\mathcal{B}(K_{L }\to \pi^{0} \nu \nu)$ & $< 6.7\times 10^{-8}$ &
 $|d_{e}|$(ecm) &$<1.6 \times 10^{-27}$ \\

$\mathcal{B}(K^{+}\to \pi^{+} \nu \nu)$ & $17.3^{+11.5}_{-10.5}\times
 10^{-11}$ & $|d_{\mu}|$(ecm) &$<2.8\times 10^{-19}$ \\

$\mathcal{B}(B_{d}\to e e)$ & $<1.13\times 10^{-7}$ &
 $|d_{\tau}|$(ecm) &$<1.1\times 10^{-17}$ \\

$\mathcal{B}(B_{d}\to \mu \mu)$ & $<0.8\times 10^{-9}$ &
 $|d_{n}|$(ecm) &$<2.9 \times 10^{-26}$ \\[1mm]

 \cline{3-4}
$\mathcal{B}(B_{d}\to \tau \tau)$ & $<4.1\times 10^{-3}$ &
 \multicolumn{2}{|c|}{$\Delta F=2$}\\
 \cline{3-4}

$\mathcal{B}(B_{s}\to e e)$ & $<7.0\times 10^{-5}$ & $|\epsilon_{K}|$
 & $(2.229 \pm 0.010)\times 10^{-3}$ \\

$\mathcal{B}(B_{s}\to \mu \mu)$ & $<4.2\times 10^{-9}$ & $\Delta
 M_{K}$ & $(5.292 \pm 0.009)\times 10^{-3}~\mathrm{ps}^{-1}$ \\

$\mathcal{B}(B_{s}\to \tau \tau)$ & $--$ & $\Delta M_{D}$ &
 $(2.37^{+0.66}_{-0.71}) \times 10^{-2}~\mathrm{ps}^{-1}$ \\

$\mathcal{B}(B_{s}\to \mu e)$ & $<2.0\times 10^{-7}$ & $\Delta
 M_{B_{d}}$ & $(0.507 \pm 0.005)~\mathrm{ps}^{-1}$ \\

$\mathcal{B}(B_{s}\to \tau e )$ & $<2.8\times 10^{-5}$ & $\Delta
 M_{B_{s}}$ & $(17.77 \pm 0.12)~\mathrm{ps}^{-1}$ \\

$\mathcal{B}(B_{s}\to \mu \tau)$ & $<2.2\times 10^{-5}$ && \\

$\mathcal{B}(B^+\to \tau^+ \nu)$ & $(1.65\pm 0.34)\times 10^{-4}$ && \\

$\frac{\mathcal{B}(B_{d}\to D\tau \nu)}{\mathcal{B}(B_{d}\to Dl \nu)}$ &
($0.407 \pm 0.12 \pm 0.049)$ && \\

$\mathcal{B}(B\to X_{s} \gamma)$ & $(3.52\pm 0.25) \times 10^{-4}$ && \\

\hline
\end{tabular}
}
\end{center}
\end{table}

\section{Application of \code{} to analysis of the $B\ra \mu^*\mu^-$ 
decay rate.}
\label{sec:bmumu}

One of the most promising signals for new physics at the LHC is the
rare decay $\bsmumu$.  It is suppressed as a loop-level
flavour-changing neutral current and by a lepton mass insertion
required for the final state muon helicities.  The LHC will be the
first experiment able to probe this decay channel down to its
SM-predicted branching ratio.  The winter 2012 experimental 95\% CL
bounds~\cite{LHCB} and the SM prediction~\cite{SMB} for $\bsmumu$
decay rate can be summarized as:\\
\begin{tabular}{p{1cm}lll}
& CMS            & $<7.7\times 10^{-9}$\\ 
& LHCb    & $<4.5\times 10^{-9}$\\
& SM Prediction  & $(3.35\pm 0.32) \times 10^{-9}$\\ 
\end{tabular}\\
ATLAS, CMS and LHCb will be able soon to reconstruct the SM-like $\bs$
signal with significance of $3\sigma$, so that this very rare decay
could be finally discovered and measured.

\code{} is an efficient tool allowing to estimate the size of possible
SUSY contributions to the $\bsmumu$ channel.   In the MSSM, even in the
restricted MFV case, for large values of $\tan\beta$ the
$\mathcal{B}(\bsmumu)$ can be strongly enhanced~\cite{BCRS} ($M_A$ is
the $CP$-odd Higgs boson mass):
\begin{eqnarray}
{\cal B}(B^0_s\rightarrow \mu^+\mu^-) &\approx& 5 \cdot 10^{-7}
\left(\frac{\tan\beta}{50}\right)^6\left(\frac{300 \hspace{.2cm}
\mathrm{GeV}}{M_A}\right)^4 \;,\label{eq:tanbeta}
\end{eqnarray}
This result can be further significantly modified by non-vanishing
flavor-violating terms in the sfermion mass matrices, leading to large
contributions from box and $Z$-penguin diagrams.   Apart from enhancing
the decay rate, the interference of these terms could also conceivably
lead to a cancellation that would suppress the branching ratio even
\textit{below} the SM prediction~\cite{DRT}.

To quantitatively study the size of possible effects, one needs to
perform a scan over the MSSM parameter space.   The ranges of variation
over MSSM parameters are shown in left panel of Table~\ref{tab:scan}
(all parameters in scan are real; ``SUSY scale'' refers to the common
mass parameter for the first two squark generations; the trilinear
soft breaking terms are set to $A_{t}=A_{b}=M_{\tilde{Q}_{L}}$ and
$A_{\tilde{\tau}}=M_{\tilde{\ell}}$).   Flavour violation is
parametrized by the ``mass insertions''~\cite{MIPORO}, where $I,J$
denote quark flavours, $X,Y$ denote superfield chirality, and $Q$
indicates the sfermion field:
\begin{eqnarray}
\delta^{IJ}_{QXY} &=& \frac{(M^2_{Q})^{IJ}_{XY}}{\sqrt{
 (M^2_{Q})^{IJ}_{XX} (M^2_{Q})^{IJ}_{YY} }}\;.  \label{eq:phys:massinsert}
\end{eqnarray}
\begin{table}[htb]
\caption{The range of input parameters and experimental constraints
  used for the numerical scan.  All mass parameters are in GeV.  }
\begin{center}
{\small
\begin{tabular}{lr}
\begin{tabular}{|lccc|}
\hline
Parameter & Min & Max & Step \\ \hline
$\tan\beta$ & 2 & 30 & varied \\
CKM phase  $\gamma$ & $0$& $\pi$ & $\pi/25$ \\
CP-odd Higgs  $M_{A}$ & 100 & 500 & 200 \\ 
Higgs mixing  $\mu$ & -450 & 450 & 300 \\ 
SU(2) wino mass  $M_{2}$ & 100 & 500 & 200 \\ 
Gluino mass  $M_{3}$ &$3M_{2}$&$3M_{2}$& 0 \\ 
SUSY scale  $M_{\mathrm{SUSY}}$ & 500 & 1000 & 500 \\ 
Slepton Masses  $M_{\tilde{\ell}}$& $\frac{M_{\mathrm{SUSY}}}{3}$ &
$\frac{M_{\mathrm{SUSY}}}{3}$ & 0 \\
Left stop   $M_{\tilde{Q}_{L}}$ & 200 & 500& 300 \\ 
Right sbottom  $M_{\tilde{b}_{R}}$ & 200& 500& 300 \\
Right stop  $M_{\tilde{t}_{R}}$ & 150 & 300 & 150 \\
$\delta_{dLL}^{13}$, $\delta_{dLL}^{23}$ & -1& 1& 0.1 \\
$\delta_{dLR}^{13}$, $\delta_{dLR}^{23}$ & -0.1& 0.1& 0.01 \\ \hline
\end{tabular} 
&
\begin{tabular}{|cc|}
\hline 
Mass & Constraint \\ \hline
$m_{\chi^0_1}$ & $>$ 46   \\
$m_{\chi^\pm_1}$ & $>$ 94   \\
$m_{\tilde b}$ &  $>$ 89  \\
$m_{\tilde t}$ & $>$ 95.7   \\
$m_{h}$ & $>$ 92.8-114 depending on $\sin^2(\alpha-\beta)$ \\
\hline 
\end{tabular}
\\
\end{tabular}
}
\label{tab:scan}
\end{center}
\end{table}

Realistic estimate of the allowed range for $\cbsmumu$ requires taking
into account the experimental constraints from measurements of other
rare decays.  For that, all $\Delta F=2$ observables,
$\mathcal{B}(B\rightarrow X_{s} \gamma), \mathcal{B}(K_{L}\rightarrow
\pi^{0} \nu \bar{\nu}), \mathcal{B}(K^+\rightarrow \pi^+ \nu
\bar{\nu})$ decay rates and the neutron and electron electric dipole
moments have been used out of quantities listed in
Table~\ref{tab:proc}.  In addition, bounds on SUSY particle masses
listed in right panel of Table~\ref{tab:scan} have been used.

For the chosen bounds from Table~\ref{tab:proc} for which the
experimental result and its error are known, parameter point in scan
was accepted if
\bea
|Q^{exp} - Q^{th}| \leq 3\Delta Q^{exp} + q |Q^{th}|.
\label{eq:xacc}
\eea
For the quantities for which only the upper bound is known,
\bea
(1+q)|Q^{th}|\leq Q^{exp}
\label{eq:xacc1}
\eea
was required.  The first and second terms on the right-hand side of
Eq.~\ref{eq:xacc} represent the $3\sigma$ experimental error and the
theoretical error respectively.  The latter differs from quantity to
quantity and is usually smaller than the value $q=50\%$ which was
assume generically to account for the limited density of a numerical
scan (see ref.~\cite{BEJR} for details).

\begin{figure}[htb]
\vspace{-3mm}
  \begin{center} \begin{tabular}{cc}
  \resizebox{75mm}{!}{\includegraphics{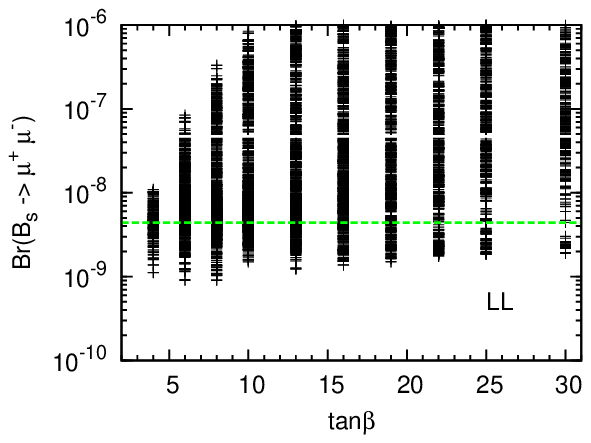}} &
  \resizebox{75mm}{!}{\includegraphics{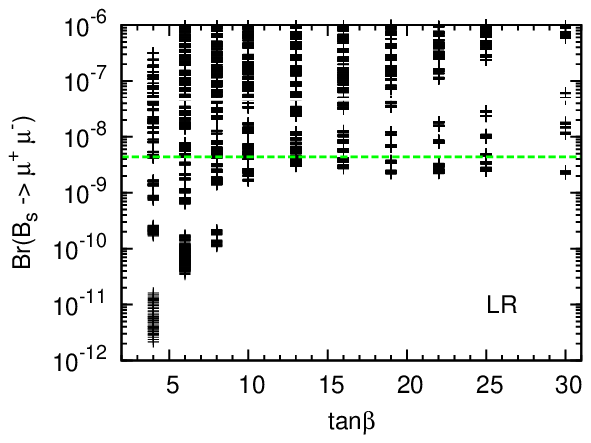}}  \\
  \end{tabular}
\vspace{-3mm}
\caption{ Predictions for ${\mathcal{B}}(B_s \to \mu^+ \mu^-)$ vs
  $\tan\beta$ from the scan of MSSM parameters in
  Table~\ref{tab:scan}. Left(right) panel: $\delta_{d \, LL}^{23}$
  ($\delta_{d \, LR}^{23}$) varied.  The dashed line shows the SM
  expectation.}
    \label{fig:scan}
  \end{center}
\end{figure}

Fig.~\ref{fig:scan} shows the predictions for $\cbsmumu$ over a
general scan of 20 million points according to Table~\ref{tab:scan}.
$\delta_{d \, LL}^{23}$ (left panel) and $\delta_{d \, LR}^{23}$
(right panel) were varied one at a time while setting the other to
zero, e.g.  all $\delta_{XY}^{{ij}}=0$ and only $\delta_{d \,
  LL}^{23}\ne 0$ in the left panel.  When $\delta_{d \, LL}^{23}$ is
varied in the range $[-1, 1]$, one finds $\cbsmumu_{min} \approx
10^{-9}$.  This minimum is almost independent of $\tan\beta$.
$|\delta_{d \, LL}^{23}|$ can take on values up to $\approx 0.9$ and
still pass all imposed constraints, though points beyond $0.3$ are
less dense.  More interesting is the case when $\delta_{d \, LR}^{23}$
is varied in the range $[-0.1,0.1]$.  One can find a narrow
cancellation region around $\delta_{d \, LR}^{23}\approx-0.01$ and
$\tan\beta \lsim 10$ where $\cbsmumu_{min} \approx 10^{-12}$.  This is
three orders of magnitude lower than the SM prediction, making it
effectively unobservable at the LHC.

\section{Conclusions}
\label{sec:conclusion}

I have presented \code, a tool capable of calculating simultaneously
29 important flavor observables in the general $R$-parity conserving
MSSM.  The calculation of the SUSY tree-level particle spectrum and
flavor mixing matrices are performed exactly.  \code{} v2 implements
also the resummation of chirally enhanced corrections, stemming from
large values of $\tan\beta$ and/or large trilinear $A$-terms.  Thus
\code{} v2 is valid for the whole parameter space of the general
$R$-parity conserving MSSM, without restrictions on the size of the
off-diagonal elements in the sfermion mass matrices - a unique feature
currently not shared by other publicly available programs calculating
FCNC and CP violation in supersymmetric models.  I hope that \code{}
becomes an important tool useful both for theorists and
experimentalists who need to perform multi-process flavor analyses
within the MSSM.

As an example of such analysis, \code{} has been used to perform a
numerical exploration of the MSSM parameter space and estimate of the
$\bsmumu$ decay rate.  Scan shows that there exist cancellation
regions where the contribution of diagrams with supersymmetric
particles interferes destructively with SM diagrams, thus allowing the
branching ratio to be significantly smaller than the SM prediction.
Such effects may effectively hide the dimuon $B^0_s$ decay mode from
the LHCb even though it is supposed to be one of the experiment's
benchmark modes.  Barring such cancellations, supersymmetric
contributions typically tend to enhance the branching ratio for
$\bsmumu$ even for moderate values of $\tan\beta\lsim 10$ so that an
experimental measurement close to the SM prediction puts strong bounds
on the size of allowed flavour violation in the squark sector.

\code{} can be downloaded from the address \webpage.



\end{document}